\documentclass{llncs}

\usepackage{listings}
\usepackage{color}
\usepackage{amssymb,amsmath,mathrsfs,stmaryrd,dsfont,bbold}
\usepackage{wrapfig}
\usepackage[lighttt]{lmodern}
\usepackage{bussproofs}
\usepackage{proof}
\usepackage{url}
\usepackage{yfonts}
\usepackage[inline]{enumitem}
\usepackage{pifont}

\lstdefinelanguage{haskell}{
  basicstyle={\ttfamily},
  columns=flexible,
  morekeywords={if, then, else, where, let, in},
  sensitive=true,
  captionpos=b
}
\lstdefinelanguage{tptp}{
  basicstyle={\ttfamily},
  columns=flexible,
  keywords={tff, type, axiom, hypothesis, conjecture},
  sensitive=true,
  captionpos=b
}

\lstdefinelanguage{cpp}{
  basicstyle={\ttfamily},
  columns=flexible,
  keywords={public, static, void, do, for, while, int, bool, if, else, break},
  sensitive=true,
  captionpos=b
}

\newcommand{\ite}[3]{\mathtt{if}\;{#1}\;\mathtt{then}\;{#2}\;\mathtt{else}\;{#3}}

\newcommand{\letin}[3]{\mathtt{let}\;{#1}\,=\,{#2}\;\mathtt{in}\;{#3}}
\newcommand{\ofsort}[2]{{{#1}:{#2}}}

\newcommand{\context}{\eta}

\newcommand{\true}{\mathit{true}}
\newcommand{\false}{\mathit{false}}
\newcommand{\bool}{\mathit{bool}}

\newcommand{\intI}{I} 

\newcommand{\interpret}[2]{\left\llbracket\,{#1}\,\right\rrbracket_{#2}}
\newcommand{\eval}[2]{\mathrm{eval}_{#2}({#1})}
\newcommand{\replacement}[3]{{#1}_{#2}^{#3}}

\newcommand{\variant}[3]{{#1}_{#2}^{#3}}

\newcommand{\folb}{{FOOL}}
\newcommand{\toFOL}[1]{\mathit{fol}(#1)}  

\newcommand{\QEDsymbol}{\text{\ding{111}}}
\newcommand{\QED}{\hfill\QEDsymbol}

\renewcommand{\implies}{\Rightarrow}
\newcommand{\liff}{\Leftrightarrow}
\newcommand{\eql}{\doteq}

\begin{document}






\title{A First Class Boolean Sort in First-Order Theorem Proving and TPTP\thanks{The final publication is available at http://link.springer.com.}}

\titlerunning{First Class Boolean Sort in TPTP}

\author{
Evgenii Kotelnikov\inst{1}
\and
Laura Kov\'acs\inst{1}%
\thanks{
The first two authors were partially supported by the Wallenberg
Academy Fellowship 2014, the Swedish VR grant D0497701, and the
Austrian research project FWF S11409-N23.}
\and
Andrei Voronkov\inst{2}%
\thanks{Partially supported by the EPSRC grant ``Reasoning in
  Verification and Security''.}
}

\institute{
Chalmers University of Technology,
Gothenburg, Sweden\\
\email{evgenyk@chalmers.se, laura.kovacs@chalmers.se}
\and
The University of Manchester,
Manchester, UK\\
\email{andrei@voronkov.com}\\
}

\authorrunning{Kotelnikov, Kov\'acs, Voronkov}

\maketitle

\begin{abstract}
To support reasoning about properties of programs operating with boolean values one needs theorem provers to be able to natively deal with the boolean sort. This way, program properties can be translated to first-order logic and theorem provers can be used to prove program properties efficiently. However, in the TPTP language, the input language of automated first-order theorem provers, the use of the boolean sort is limited compared to other sorts, thus hindering the use of first-order theorem provers in program analysis and verification. In this paper, we present an extension \folb\ of many-sorted first-order logic, in which the boolean sort is treated as a first-class sort. Boolean terms are indistinguishable from formulas and can appear as arguments to functions. In addition, \folb\ contains \verb'if'-\verb'then'-\verb'else' and \verb'let'-\verb'in' constructs. We define the syntax and semantics of \folb\ and its model-preserving translation to first-order logic. We also introduce a new technique of dealing with boolean sorts in superposition-based theorem provers. Finally, we discuss how the TPTP language can be changed to support \folb.
\end{abstract}




\section{Introduction}
\label{sec:introduction}


Automated program analysis and verification requires 
discovering and proving program properties. Typical examples of such properties are loop invariants or Craig interpolants. These properties usually are expressed in combined theories of various data structures, such as integers and arrays, and hence require reasoning with both theories and quantifiers. Recent approaches in interpolation and loop invariant generation~\cite{McMillan08,FaseKV09,hoder2012popl} present initial results of using first-order theorem provers for generating quantified program properties. First-order theorem provers can also be used to generate program properties with quantifier alternations~\cite{FaseKV09}; such properties could not be generated fully automatically by any previously known method. 
Using first-order theorem prover to generate, and not only prove program properties, opens new directions in analysis and verification of real-life programs. 

First-order theorem provers, such as iProver~\cite{iProver}, E~\cite{E13}, and Vampire~\cite{Vampire13}, lack however various features that are crucial for program analysis. For example, first-order theorem provers do not yet efficiently handle (combinations of) theories;
nevertheless, sound but incomplete theory axiomatisations can be used in a first-order prover even for theories having no finite axiomatisation. Another difficulty in modelling properties arising in program analysis using theorem provers is the gap between the semantics of expressions used in programming languages and expressiveness of the logic used by the theorem prover. A similar gap exists between the language used in presenting mathematics. For example, a standard way to capture assignment in program analysis is to use a \verb'let'-\verb'in' expression, which introduces a local binding of a variable, or a function for array assignments, to a value. There is no local binding expression in first-order logic, which means that any modelling of imperative programs using first-order theorem provers at the backend, should implement a translation of \verb'let'-\verb'in' expressions. Similarly, mathematicians commonly use local definitions within definitions and proofs. Some functional programming languages also contain expressions introducing local bindings. In all three cases, to facilitate the use of first-order provers, one needs a theorem prover implementing \verb'let'-\verb'in' constructs natively.

Efficiency of reasoning-based program analysis largely depends on how programs are translated into a collection of logical formulas capturing the program semantics. The boolean structure of a program property that can be efficiently treated by a theorem prover is however very sensitive to the architecture of the reasoning engine of the prover. Deriving and expressing program properties in the ``right'' format therefore requires solid knowledge about how theorem provers work and are implemented~--- something that a user of a verification tool might not have. Moreover, it can be hard to efficiently reason about certain classes of program properties, unless special inference rules and heuristics are added to the theorem prover, see e.g.~\cite{ATVA14} when it comes to prove properties of data collections with extensionality axioms. 

In order to increase the expressiveness of program properties generated by reasoning-based program analysis, the language of logical formulas accepted by a theorem prover needs to be extended with constructs of programming languages. This way, a straightforward translation of programs into first-order logic can be achieved, thus relieving users from designing translations which can be efficiently treated by the theorem prover. 
One example of such an extension is recently added to the TPTP language~\cite{TPTP} of first-order theorem provers, resembling \verb'if'-\verb'then'-\verb'else' and \verb'let'-\verb'in' expressions that are common in programming languages. Namely, special functions \verb'$ite_t' and \verb'$ite_f' can respectively be used to express a conditional statement on the level of logical terms and formulas, and \verb'$let_tt', \verb'$let_tf', \verb'$let_ff' and \texttt{\$let\_ft} can be used to express local variable bindings for all four possible combinations of logical terms (\verb't') and formulas (\verb'f'). While satisfiability modulo theory (SMT) solvers, such as Z3~\cite{Z3} and CVC4~\cite{CVC4}, integrate \verb'if'-\verb'then'-\verb'else' and \verb'let'-\verb'in' expressions, in the first-order theorem proving community so far only Vampire supports such expressions.

To illustrate the advantage of using \verb'if'-\verb'then'-\verb'else' and \verb'let'-\verb'in' expressions in automated provers, let us consider the following  simple example. We are interested in verifying the partial correctness of the code fragment below:
\begin{lstlisting}[language=haskell]
if (r(a)) {
  a := a + 1
} else {
  a := a + q(a)
}
\end{lstlisting}
using the pre-condition $((\forall x) P(x) \Rightarrow x \ge 0) \wedge ((\forall x) \mathtt{q}(x) > 0) \wedge P(\mathtt{a})$ and the post-condition $\mathtt{a} > 0$.
Let $\mathtt{a1}$ denote the value of the program variable $\mathtt{a}$ after the execution of the \verb'if'-statement. Using \verb'if'-\verb'then'-\verb'else' and \verb'let'-\verb'in' expressions, the next state function for $\mathtt{a}$ can naturally be expressed by the following formula: 
\begin{lstlisting}[language=haskell]
a1 = if r(a) then let a = a + 1 in a
               else let a = a + q(a) in a
\end{lstlisting}

This formula can further be encoded in TPTP, and hence used by a theorem prover as a hypothesis in proving partial correctness of the above code snippet. We illustrate below the TPTP encoding of the first-order problem corresponding to the partial program correctness problem we consider.  Note that the pre-condition becomes a hypothesis in TPTP, whereas the proof obligation given by the post-condition is a TPTP conjecture. All formulas below are typed first-order formulas (\verb'tff') in TPTP that use the built-in integer sort (\verb'$int'). 
{\small\begin{lstlisting}[language=tptp]
tff(1, type, p : $int > $o).
tff(2, type, q : $int > $int).
tff(3, type, r : $int > $o).
tff(4, type, a : $int).
tff(5, hypothesis, ! [X : $int] : (p(X) => $greatereq(X, 0))).
tff(6, hypothesis, ! [X : $int] : ($greatereq(q(X), 0))).
tff(7, hypothesis, p(a)).
tff(8, hypothesis,
    a1 = $ite_t(r(a), $let_tt(a, $sum(a, 1), a),
                        $let_tt(a, $sum(a, q(a)), a))).
tff(9, conjecture, $greater(a1, 0)). 
\end{lstlisting}}

Running a theorem prover that supports \verb'$ite_t' and \verb'$let_tt' on this TPTP problem would prove the partial correctness of the program we considered. Note that without the use of \verb'if'-\verb'then'-\verb'else' and \verb'let'-\verb'in' expressions, a more tedious translation is needed for expressing the next state function of the program variable $\mathtt{a}$ as a first-order formula. When considering more complex programs containing multiple conditional expressions assignments and composition, 
computing the next state function of a program variable results in a formula of size exponential in the number of conditional expressions. This problem of computing the next state function of variables is well-known in the program analysis community, by computing so-called static single assignment (SSA) forms. Using the \verb'if'-\verb'then'-\verb'else' and \verb'let'-\verb'in' expressions recently introduced in TPTP and already implemented in Vampire \cite{PSI14}, one can have a linear-size translation instead.

Let us however note that the usage of conditional expressions in TPTP is somewhat limited. The first argument of \verb'$ite_t' and \verb'$ite_f' is a logical formula, which means that a boolean condition from the program definition should be translated as such. At the same time, the same condition can be treated as a value in the program, for example, in a form of a boolean flag, passed as an argument to a function. Yet we cannot mix terms and formulas in the same way in a logical statement. 
A possible solution would be to map the boolean type of programs to a user-defined boolean sort, postulate axioms about its semantics, and manually convert boolean terms into formulas where needed. This approach, however, suffers the disadvantages mentioned earlier, namely the need to design a special translation and its possible inefficiency.

Handling boolean terms as formulas is needed not only in applications of reasoning-based program analysis, but also in various problems of formalisation of mathematics. 
For example, if one looks at two largest kinds of attempts to formalise mathematics and proofs: those performed by interactive proof assistants, such as Isabelle~\cite{Isabelle},  and the Mizar project~\cite{Mizar}, one can see that first-order theorem provers are the main workhorses behind computer proofs in both cases -- see e.g.~\cite{Sledgehammer,DBLP:conf/icms/UrbanHV10}. 
Interactive theorem provers, such as Isabelle routinely use quantifiers over booleans.  Let us illustrate this by the 
following examples, chosen among 490 properties about (co)algebraic datatypes, featuring quantifiers over booleans, generated by Isabelle and kindly found for us by Jasmin Blanchette. Consider the distributivity of a conditional expression (denoted by the $ite$ function) over logical connectives, a pattern that is
widely used in reasoning about properties of data structures. For lists and the \verb'contains' function that checks that its second argument contains the first one, we have the following example: 
\begin{gather}\label{formula:contains}
  \begin{aligned}
&(\forall\ofsort{p}{\bool})(\forall\ofsort{l}{list_A})(\forall\ofsort{x}{A})(\forall\ofsort{y}{A}) \\
&\hspace{0.4em}\mathtt{contains}(l,\mathrm{ite}(p,\,x,\,y)) \doteq \\
&\hspace{2em} (p \Rightarrow \mathtt{contains}(l,\,x)) \wedge (\neg p \Rightarrow \mathtt{contains}(l,\,y))
 \end{aligned}
\end{gather}

A more complex example with a heavy use of booleans is the unsatisfiability of the definition of \verb'subset_sorted'. The \verb'subset_sorted' function takes two sorted lists and checks that its second argument is a sublist of the first one.
\begin{equation}\label{formula:subset-sorted}
\begin{aligned}
&(\forall\ofsort{l_1}{list_A})(\forall\ofsort{l_2}{list_A})(\forall\ofsort{p}{Bool}) \\
&\hspace{0.5em}\neg (\mathtt{subset\_sorted}(l_1,\,l_2) \doteq p ~\wedge \\
&\hspace{1.6em}      (\forall\ofsort{l_2'}{list_A})\neg (l_1 \doteq \mathtt{nil} \wedge l_2 \doteq l_2' \wedge p) ~\wedge \\
&\hspace{1.6em}      (\forall\ofsort{x_1}{A})(\forall\ofsort{l_1'}{list_A})\neg (l_1 \doteq \mathtt{cons}(x_1,\,l_1') \wedge l_2 \doteq \mathtt{nil} \wedge \neg p) ~\wedge \\
&\hspace{1.6em}      (\forall\ofsort{x_1}{A})(\forall\ofsort{l_1'}{list_A})(\forall\ofsort{x_2}{A})(\forall\ofsort{l_2'}{list_A}) \\
&\hspace{2.1em}       \neg (l_1 \doteq \mathtt{cons}(x_1,\,l_1') \wedge l_2 \doteq \mathtt{cons}(x_2,\,l_2') ~\wedge \\
&\hspace{3.3em}       p \doteq \mathrm{ite}(x_1 < x_2,\,\false,\\
&\hspace{10.7em}                       \mathrm{ite}(x_1 \doteq x_2,\,\mathtt{subset\_sorted}(l_1',\,l_2'), \\
&\hspace{16.25em}                                                                      \mathtt{subset\_sorted}(\mathtt{cons}(x_1,\,l_1'),\,l_2')))))
\end{aligned}
\end{equation}

Formulas with boolean terms are also common in the SMT-LIB project~\cite{SMT-LIB}, the collection of benchmarks for SMT-solvers. Its core logic is a variant of first-order logic that treats boolean terms as formulas, in which logical connectives and conditional expressions are defined in the core theory.


In this paper we propose a modification \folb\ of first-order logic, which includes a first-class boolean sort and \verb'if'-\verb'then'-\verb'else' and \verb'let'-\verb'in' expressions, aimed for being used in automated first-order theorem proving. It is the smallest logic that contains both the SMT-LIB core theory and the monomorphic first-order subset of TPTP. The syntax and semantics of the logic are given in Section~\ref{sec:folbool}. 
We further describe how \folb\ can be translated to the ordinary many-sorted first-order logic in Section~\ref{sec:folb-to-fol}.
Section~\ref{sec:superposition} discusses superposition-based theorem proving and proposes a new way of dealing with the boolean sort in it. 
In Section~\ref{sec:tptp} we discuss the support of the boolean sort in TPTP and propose changes to it required to support a first-class boolean sort. We point out that such changes can also partially simplify the syntax of TPTP. 
Section~\ref{sec:related} discusses related work and Section~\ref{sec:conclusions} contains concluding remarks.

The main contributions of this paper are the following:

\begin{enumerate}
\item the definition of \folb\ and its semantics;
\item a translation from \folb\ to first-order logic, which can be used to support \folb\ in existing first-order theorem provers;
\item a new technique of dealing with the boolean sort in superposition theorem provers, allowing one to replace boolean sort axioms by special rules;
\item a proposal of a change to the TPTP language, intended to support \folb\ and also simplify \verb'if'-\verb'then'-\verb'else' and \verb'let'-\verb'in' expressions.
\end{enumerate}

\section{First-Order Logic with Boolean Sort}
\label{sec:folbool}

First-order logic with the boolean sort (\folb) extends many-sorted first-order logic (FOL) in two ways:
\begin{enumerate}
\item formulas can be treated as terms of the built-in boolean sort; and
\item one can use \verb'if'-\verb'then'-\verb'else' and \verb'let'-\verb'in' expressions defined below.
\end{enumerate}
\folb\ is the smallest logic containing both the SMT-LIB core theory and the monomorphic first-order part of the TPTP language. It extends the SMT-LIB core theory by adding \texttt{let-in} expressions defining functions and TPTP by the first-class boolean sort.

\subsection{Syntax}

We assume a countable infinite set of \emph{variables}.

\begin{definition}\label{def:folb-signature}\em
  A \emph{signature} of first-order logic with the boolean sort is a triple $\Sigma = (S, F, \context)$, where:

  \begin{enumerate}
  \item $S$ is a set of \emph{sorts}, which contains a special sort $\bool$. A \emph{type} is either a sort or a non-empty sequence $\sigma_1,\ldots,\sigma_n,\sigma$ of sorts, written as $\sigma_1 \times \ldots \times \sigma_n \to \sigma$. When $n = 0$, we will simply write $\sigma$ instead of $\to\sigma$. We call a \emph{type assignment} a mapping from a set of variables and function symbols to types, which maps variables to sorts.

    \item $F$ is a set of \emph{function symbols}. We require $F$ to contain binary function symbols $\vee$, $\wedge$, $\implies$ and $\liff$, used in infix form, a unary function symbol $\neg$, used in prefix form, and nullary function symbols $\true$, $\false$.

    \item $\context$ is a \emph{type assignment} which maps each function symbol $f$ into a type $\tau$. When the signature is clear from the context, we will write $\ofsort{f}{\tau}$ instead of $\context(f)=\tau$ and say that $f$ is of the type $\tau$.

    We require the symbols $\vee, \wedge, \implies, \liff$ to be of the type $\bool \times \bool \to \bool$, $\neg$ to be of the type $\bool \to \bool$ and $\true,\false$ to be of the type $\bool$. \QED
  \end{enumerate}
\end{definition}
In the sequel we assume that $\Sigma = (S,F,\context)$ is an arbitrary but fixed signature.

To define the semantics \folb, we will have to extend the signature and also assign sorts to variables. Given a type assignment $\context$, we define $\context,x:\sigma$ to be the type assignment that maps a variable $x$ to $\sigma$ and coincides otherwise with $\context$. Likewise, we define $\context,f:\tau$ to be the type assignment that maps a function symbol $f$ to $\tau$ and coincides otherwise with $\context$.

Our next aim to define the set of terms and their sorts with respect to a type assignment $\context$. This will be done using a relation $\context \vdash t:\sigma$, where $\sigma \in S$, terms can then be defined as all such expressions $t$.

\begin{definition}\label{def:folb-terms}\rm
  The relation $\context \vdash t:\sigma$, where $t$ is an expression and $\sigma \in S$ is defined inductively as follows. If $\context \vdash t:\sigma$, then we will say that $t$ is a \emph{term of the sort $\sigma$} w.r.t.\ $\context$. 
  \begin{enumerate}
    \item If $\context(x) = \sigma$, then $\context \vdash x:\sigma$.

    \item If $\context(f) = \sigma_1 \times \ldots \times \sigma_n \to \sigma$, $\context \vdash t_1:\sigma_1$, \ldots, $\context \vdash t_n:\sigma_n$, then $\context \vdash  f(t_1, \ldots, t_n) : \sigma$.

    \item If $\context \vdash \phi:\bool$, $\context \vdash t_1:\sigma$ and $\context \vdash t_2:\sigma$, then $\context \vdash (\ite{\phi}{t_1}{t_2}) : \sigma$.

    \item Let $f$ be a function symbol and $x_1,\ldots,x_n$ pairwise distinct variables. If $\context,x_1:\sigma_1,\ldots,x_n:\sigma_n \vdash s:\sigma$ and $\context,f:(\sigma_1\times \ldots \times\sigma_n \to\sigma) \vdash t : \tau$, then $\context \vdash (\letin{f(x_1:\sigma_1, \ldots, x_n:\sigma_n)}{s}{t}) : \tau$.

    \item If $\context \vdash  s:\sigma$ and $\context \vdash  t:\sigma$, then $\context \vdash (s \eql t) : \bool$.

    \item If $\context,x : \sigma \vdash \phi : \bool$, then $\context \vdash (\forall x : \sigma)\phi : \bool$ and $\context \vdash (\exists x:\sigma)\phi : \bool$. \QED
  \end{enumerate}
\end{definition}
We only defined a \verb'let'-\verb'in' expression for a single function symbol. It is not hard to extend it to a \verb'let'-\verb'in' expression that binds multiple pairwise distinct function symbols in parallel, the details of such an extension are straightforward.

When $\context$ is the type assignment function of $\Sigma$ and $\context \vdash t : \sigma$, we will say that $t$ is a \emph{$\Sigma$-term of the sort $\sigma$}, or simply that $t$ is \emph{a term of the sort $\sigma$}. It is not hard to argue that every $\Sigma$-term has a unique sort.

According to our definition, not every term-like expression has a sort. For example, if $x$ is a variable and $\context$ is not defined on $x$, then $x$ is a not a $term$ w.r.t.\ $\context$. To make the relation between term-like expressions and terms clear, we introduce a notion of free and bound occurrences of variables and function symbols. We call the following occurrences of variables and function symbols \emph{bound}:

\begin{enumerate}
\item any occurrence of $x$ in $(\forall x:\sigma) \phi$ or in $(\exists x:\sigma) \phi$;
\item in the term $\letin{f(x_1:\sigma_1, \ldots, x_n:\sigma_n)}{s}{t}$ any occurrence of a variable $x_i$ in $f(x_1:\sigma_1, \ldots, x_n:\sigma_n)$ or in $s$, where $i = 1,\ldots, n$.
\item in the term $\letin{f(x_1:\sigma_1, \ldots, x_n:\sigma_n)}{s}{t}$ any occurrence of the function symbol $f$ in $f(x_1:\sigma_1, \ldots, x_n:\sigma_n)$ or in $t$.
\end{enumerate}
All other occurrences are called \emph{free}. We say that a variable or a function symbol is \emph{free} in a term $t$ if it has at least one free occurrence in $t$. A term is called \emph{closed} if it has no occurrences of free variables.

\begin{theorem}\rm
  Suppose $\context \vdash t : \sigma$. Then
  \begin{enumerate}
    \item for every free variable $x$ of $t$, $\context$ is defined on $x$;
    \item for every free function symbol $f$ of $t$, $\context$ is defined on $f$;
    \item if $x$ is a variable not free in $t$, and $\sigma'$ is an arbitrary sort, then
      $\context, x : \sigma' \vdash t : \sigma$;
    \item if $f$ is a function symbol not free in $t$, and $\tau$ is an arbitrary type, then $\context, f : \tau \vdash t : \sigma$. \QED
  \end{enumerate}
\end{theorem}

\begin{definition}\rm
  A \emph{predicate symbol} is any function symbol of the type $\sigma_1 \times \ldots \times \sigma_n \to \bool$. 
  A \emph{$\Sigma$-formula} is a $\Sigma$-term of the sort $\bool$. All $\Sigma$-terms that are not $\Sigma$-formulas are called \emph{non-boolean terms}. \QED
\end{definition}

Note that, in addition to the use of \texttt{let-in} and \texttt{if-then-else}, \folb\ is a proper extension of first-order logic. For example, in \folb\ formulas can be used as arguments to terms and one can quantify over booleans. As a consequence, every quantified boolean formula is a formula in \folb.

\subsection{Semantics}

As usual, the semantics of \folb\ is defined by introducing a notion of \emph{interpretation} and defining how a term is evaluated in an interpretation.

\begin{definition}\label{def:folb-interpretation}\rm
  Let $\context$ be a type assignment. 
  A \emph{$\context$-interpretation} $\intI$ is a map, defined as follows. Instead of $\intI(e)$ we will write $\interpret{e}{\intI}$, for every element $e$ in the domain of $\intI$.
  \begin{enumerate}
    \item Each sort $\sigma \in S$ is mapped to a nonempty domain $\interpret{\sigma}{\intI}$. We require $\interpret{\bool}{\intI} = \left\{0, 1\right\}$.

    \item If $\context \vdash x:\sigma$, then $\interpret{x}{\intI} \in \interpret{\sigma}{\intI}$.

    \item If $\context(f) = \sigma_1 \times \ldots \times \sigma_n \to \sigma$, then $\interpret{f}{\intI}$ is a function from $\interpret{\sigma_1}{\intI} \times \ldots \times \interpret{\sigma_n}{\intI}$ to $\interpret{\sigma}{\intI}$.

    \item We require $\interpret{\true}{\intI} = 1$ and $\interpret{\false}{\intI} = 0$. We require $\interpret{\wedge}{\intI}$, $\interpret{\vee}{\intI}$, $\interpret{\implies}{\intI}$, $\interpret{\liff}{\intI}$ and $\interpret{\neg}{\intI}$ respectively to be the logical conjunction, disjunction, implication, equivalence and negation, defined over $\{0,1\}$ in the standard way. 
  \end{enumerate}
\end{definition}

Given a $\context$-interpretation $\intI$ and a function symbol $f$, we define $\variant{\intI}{f}{g}$ to be the mapping that maps $f$ to $g$ and coincides otherwise with $\intI$. 
Likewise, for a variable $x$ and value $a$ we define $\variant{\intI}{x}{a}$ to be the mapping that maps $x$ to $a$ and coincides otherwise with $\intI$. 

\begin{definition}\label{def:folb-term-evaluation}\rm
  Let $\intI$ be a $\context$-interpretation, and $\context \vdash t:\sigma$. The \emph{value of $t$ in $\intI$}, denoted as $\eval{t}{\intI}$, is a value in $\interpret{\sigma}{\intI}$ inductively defined as follows:

  \[
    \begin{array}{rcl}
      \eval{x}{\intI} & = & \interpret{x}{\intI}.
      \\*[1ex]
      \eval{f(t_1, \ldots, t_n)}{\intI} & = &
        \interpret{f}{\intI}(\eval{t_1}{\intI}, \ldots, \eval{t_n}{\intI}).
      \\*[1ex]
      \eval{\ite{\phi}{s}{t}}{\intI} & = &
        \left\{ \begin{array}{ll}
                  \eval{s}{\intI}, & \text{if $\eval{\phi}{\intI} = 1$;} \\*[1ex]
                  \eval{t}{\intI}, & \text{otherwise.}
                \end{array}\right.
      \\*[1ex]
      \eval{\letin{f(x_1:\sigma_1,\ldots,x_n:\sigma_n)}{s}{t}}{\intI} & = &
        \eval{t}{\replacement{\intI}{f}{g}},
    \end{array}
  \]
  where $g$ is such that for all $i = 1, \ldots, n$ and $a_i \in \interpret{\sigma_i}{\intI}$, we have $g(a_1, \ldots, a_n) = \eval{s}{\replacement{\intI}{x_1 \ldots x_n}{a_1 \ldots a_n}}$.
  \[
    \begin{array}{rcl}
      \eval{s \eql t}{\intI} & = &
        \left\{ \begin{array}{ll}
                  1, & \text{if } \eval{s}{\intI} = \eval{t}{\intI}; \\*[1ex]
                  0, & \text{otherwise.}
                \end{array}\right.
      \\*[1ex]
      \eval{(\forall x : \sigma)\phi}{\intI} & = &
        \left\{ \begin{array}{ll}
                  1, & \text{if } \eval{\phi}{\replacement{\intI}{x}{a}} = 1\\*[1ex]
                     & \text{~~~for all } a \in \intI(\sigma); \\*[1ex]
                  0, & \text{otherwise.}
                \end{array}\right.
      \\*[1ex]
      \eval{(\exists x : \sigma)\phi}{\intI} & = &
        \left\{ \begin{array}{ll}
                  1, & \text{if } \eval{\phi}{\replacement{\intI}{x}{a}} = 1\\*[1ex]
                     & \text{~~~for some } a \in \intI(\sigma); \\*[1ex]
                  0, & \text{otherwise.}
                \end{array}\right.
    \end{array}
  \]
\end{definition}

\begin{theorem}\label{thm:semantics}\rm
  Let $\context \vdash \phi : \bool$ and $\intI$ be a $\context$-interpretation. Then
  \begin{enumerate}
    \item for every free variable $x$ of $\phi$, $\intI$ is defined on $x$;
    \item for every free function symbol $f$ of $\phi$, $\intI$ is defined on $f$;
    \item if $x$ is a variable not free in $\phi$, $\sigma$ is an arbitrary sort, and $a \in \interpret{\sigma}{\intI}$ then $\eval{\phi}{\intI} = \eval{\phi}{\replacement{\intI}{x}{a}}$;
    \item if $f$ is a function symbol not free in $\phi$, $\sigma_1,\ldots,\sigma_n,\sigma$ are arbitrary sorts and $g \in \interpret{\sigma_1}{\intI} \times \ldots \times \interpret{\sigma_n}{\intI} \to \interpret{\sigma}{\intI}$, then $\eval{\phi}{\intI} = \eval{\phi}{\replacement{\intI}{f}{g}}$. \QED
  \end{enumerate}
\end{theorem}

Let $\context \vdash \phi : \bool$. A $\context$-interpretation $\intI$ is called a \emph{model} of $\phi$, denoted by $\intI \models \phi$, if $\eval{\phi}{\intI} = 1$. If $\intI \models \phi$, we also say that $\intI$ \emph{satisfies} $\phi$. We say that $\phi$ is \emph{valid}, if $\intI \models \phi$ for all $\context$-interpretations $\intI$, and \emph{satisfiable}, if $\intI \models \phi$ for at least one $\context$-interpretation $\intI$. Note that Theorem~\ref{thm:semantics} implies that any interpretation, which coincides with $\intI$ on free variables and free function symbols of $\phi$ is also a model of $\phi$.

\section{Translation of \folb{} to FOL}
\label{sec:folb-to-fol}

\folb{} is a modification of FOL. Every FOL formula is syntactically a \folb{} formula and has the same models, but not the other way around. In this section we present a translation from \folb{} to FOL, which preserves models of $\phi$. This translation can be used for proving theorems of \folb\ using a first-order theorem prover. We do not claim that this translation is efficient -- more research is required on designing translations friendly for first-order theorem provers.

We do not formally define many-sorted FOL with equality here, since FOL is essentially a subset of \folb, which we will discuss now.  

We say that an occurrence of a subterm $s$ of the sort $\bool$ in a term $t$ is in a \emph{formula context} if it is an argument of a logical connective or the occurrence in either $(\forall x:\sigma)s$ or $(\exists x:\sigma)s$. We say that an occurrence of $s$ in $t$ is in a \emph{term context} if this occurrence is an argument of a function symbol, different from a logical connective, or an equality. We say that a formula of \folb{} is \emph{syntactically first order} if it contains no \verb'if'-\verb'then'-\verb'else' and \verb'let'-\verb'in' expressions, no variables occurring in a formula context and no formulas occurring in a term context. By restricting the definition of terms to the subset of syntactically first-order formulas, we obtain the standard definition of many-sorted first-order logic, with the only exception of having a distinguished boolean sort and constants $\true$ and $\false$ occurring in a formula context.

Let $\phi$ be a closed $\Sigma$-formula of \folb{}. We will perform the following steps to translate $\phi$ into a first-order formula. During the translation we will maintain a set of formulas $D$, which initially is empty. The purpose of $D$ is to collect a set of formulas (definitions of new symbols), which guarantee that the transformation preserves models.

\begin{enumerate}
\item Make a sequence of translation steps obtaining a syntactically first order formula $\phi'$. During this translation we will introduce new function symbols and add their types to the type assignment $\context$. We will also add formulas describing properties of these symbols to $D$. The translation will guarantee that the formulas $\phi$ and $\bigwedge_{\psi \in D}\psi \wedge \phi'$ are equivalent, that is, have the same models restricted to $\Sigma$.

\item Replace the constants $\true$ and $\false$, standing in a formula context, by nullary predicates $\top$ and $\bot$ respectively, obtaining a first-order formula.

\item Add special boolean sort axioms.
\end{enumerate}
During the translation, we will say that a function symbol or a variable is \emph{fresh} if it neither appears in $\phi$ nor in any of the definitions, nor in the domain of $\context$.

We also need the following definition. Let $\context \vdash t:\sigma$, and $x$ be a variable occurrence in $t$. The \emph{sort of this occurrence of $x$} is defined as follows:

\begin{enumerate}
\item any free occurrence of $x$ in a subterm $s$ in the scope of $(\forall x:\sigma')s$ or $(\exists x:\sigma')s$ has the sort $\sigma'$.
\item any free occurrence of $x_i$ in a subterm $s_1$ in the scope of \\$\letin{f(x_1:\sigma_1, \ldots, x_n:\sigma_n)}{s_1}{s_2}$ has the sort $\sigma_i$, where $i = 1,\ldots,n$. 
\item a free occurrence of $x$ in $t$ has the sort $\context(x)$.
\end{enumerate}
If $\context \vdash t:\sigma$, $s$ is a subterm of $t$ and $x$ a free variable in $s$, we say that $x$ has a sort $\sigma'$ in $s$ if its free occurrences in $s$ have this sort.

The translation steps are defined below. We start with an empty set $D$ and an initial \folb\ formula $\phi$, which we would like to change into a syntactically first-order formula. At every translation step we will select a formula $\chi$, which is either $\phi$ or a formula in $D$, which is not syntactically first-order, replace a subterm in $\chi$ it by another subterm, and maybe add a formula to $D$. The translation steps can be applied in any order.

\begin{enumerate}
  \item Replace a boolean variable $x$ occurring in a formula context, by $x \eql \true$.

  \item Suppose that $\psi$ is a formula occurring in a term context such that (i) $\psi$ is different from $\true$ and $\false$, (ii) $\psi$ is not a variable, and (iii) $\psi$ contains no free occurrences of function symbols bound in $\chi$. Let $x_1,\ldots,x_n$ be all free variables of $\psi$ and $\sigma_1,\ldots,\sigma_n$ be their sorts. Take a fresh function symbol $g$, add the formula $(\forall x_1:\sigma_1)\ldots(\forall x_n:\sigma_n) (\psi \liff g(x_1,\ldots,x_n) \eql \true)$ to $D$ and replace $\psi$ by $g(x_1,\ldots,x_n)$. Finally, change $\context$ to $\context,g : \sigma_1 \times \ldots \times \sigma_n \to \bool$.

  \item Suppose that $\ite{\psi}{s}{t}$ is a term containing no free occurrences of function symbols bound in $\chi$. Let $x_1,\ldots,x_n$ be all free variables of this term and $\sigma_1,\ldots,\sigma_n$ be their sorts. Take a fresh function symbol $g$, add the formulas $(\forall x_1:\sigma_1)\ldots(\forall x_n:\sigma_n) (\psi \implies g(x_1,\ldots,x_n) \eql s)$ and $(\forall x_1:\sigma_1)\ldots(\forall x_n:\sigma_n) (\neg\psi \implies g(x_1,\ldots,x_n) \eql t)$ to $D$ and replace this term by $g(x_1,\ldots,x_n)$. Finally, change $\context$ to $\context,g : \sigma_1 \times \ldots \times \sigma_n \to \sigma_0$, where $\sigma_0$ is such that $\context,x_1:\sigma_1,\ldots,x_n:\sigma_n \vdash s : \sigma_0$.

  \item Suppose that $\letin{f(x_1:\sigma_1, \ldots, x_n:\sigma_n)}{s}{t}$ is a term containing no free occurrences of function symbols bound in $\chi$. Let $y_1,\ldots,y_m$ be all free variables of this term and $\tau_1,\ldots,\tau_m$ be their sorts. Note that the variables in $x_1,\ldots,x_n$ are not necessarily disjoint from the variables in $y_1,\ldots,y_m$. 

Take a fresh function symbol $g$ and fresh sequence of variables $z_1,\ldots,z_n$. Let the term $s'$ be obtained from $s$ by replacing all free occurrences of $x_1,\ldots,x_n$ by $z_1,\ldots,z_n$, respectively. Add the formula $(\forall z_1:\sigma_1)\ldots(\forall z_n:\sigma_n) (\forall y_1:\tau_1)\ldots(\forall y_m:\tau_m) (g(z_1,\ldots,z_n,y_1,\ldots,y_m) \eql s')$ to $D$. Let the term $t'$ be obtained from $t$ by replacing all bound occurrences of $y_1,\ldots,y_m$ by fresh variables and each application $f(t_1, \ldots, t_n)$ of a free occurrence of $f$ in $t$ by $g(t_1, \ldots, t_n,\allowbreak y_1, \ldots, y_m)$. Then replace $\letin{f(x_1:\sigma_1, \ldots, x_n:\sigma_n)}{s}{t}$ by $t'$. Finally, change $\context$ to $\context,g : \sigma_1 \times \ldots \times \sigma_n \times \tau_1 \times \ldots \times \tau_m \to \sigma_0$, where $\sigma_0$ is such that $\context,x_1:\sigma_1,\ldots,x_n:\sigma_n,y_1:\tau_1,\ldots,y_m:\tau_m \vdash s : \sigma_0$. 
\end{enumerate}
The translation terminates when none of the above rules apply.

We will now formulate several of properties of this translation, which will imply that, in a way, it preserves models. These properties are not hard to prove, we do not include proofs in this paper.

\begin{lemma}\label{lemma:step-preserves-equivalence}\rm
  Suppose that a single step of the translation changes a formula $\phi_1$ into $\phi_2$, $\delta$ is the formula added at this step (for step 1 we can assume $\true=\true$ is added), $\context$ is the type assignment before this step and $\context'$ is the type assignment after. Then for every $\context'$-interpretation $\intI$ we have $\intI \models \delta \implies (\phi_1 \liff \phi_2)$. \QED
\end{lemma}

By repeated applications of this lemma we obtain the following result.

\begin{lemma}\label{lemma:definitions-preserve-models}\rm
  Suppose that the translation above changes a formula $\phi$ into $\phi'$, $D$ is the set of definitions obtained during the translation, $\context$ is the initial type assignment and $\context'$ is the final type assignment of the translation. Let $I'$ be any interpretation of $\context'$. Then $I' \models \bigwedge_{\psi \in D} \psi \implies (\phi \Leftrightarrow \phi')$. \QED
\end{lemma}

We also need the following result.

\begin{lemma}\label{lem:termination}\rm
  Any sequence of applications of the translation rules terminates. \QED
\end{lemma}

The lemmas proved so far imply that the translation terminates and the final formula is equivalent to the initial formula in every interpretation satisfying all definitions in $D$. To prove model preservation, we also need to prove some properties of the introduced definitions. 

\begin{lemma}\label{lem:satisfy:definitions}\rm
  Suppose that one of the steps 2--4 of the translation translates a formula $\phi_1$ into $\phi_2$, $\delta$ is the formula added at this step, $\context$ is the type assignment before this step, $\context'$ is the type assignment after, and $g$ is the fresh function symbol introduced at this step. Let also $\intI$ be $\context$-interpretation. Then there exists a function $h$ such that $\replacement{\intI}{g}{h} \models \delta$. \QED
\end{lemma}

These properties imply the following result on model preservation.

\begin{theorem}\label{thm:model:preservation}\rm
  Suppose that the translation above translates a formula $\phi$ into $\phi'$, $D$ is the set of definitions obtained during the translation, $\context$ is the initial type assignment and $\context'$ is the final type assignment of the translation. 
  \begin{enumerate}
    \item Let $\intI$ be any $\context$-interpretation. Then there is a $\context'$-interpretation $I'$ such that $\intI'$ is an extension of $\intI$ and $\intI' \models \bigwedge_{\psi \in D} \psi \wedge \phi'$.
    \item Let $\intI'$ be a $\context'$-interpretation and $\intI' \models \bigwedge_{\psi \in D} \psi \wedge \phi'$. Then $\intI' \models \phi$. \QED
  \end{enumerate}
\end{theorem}
This theorem implies that $\phi$ and $\bigwedge_{\psi \in D} \psi \wedge \phi'$ have the same models, as far as the original type assignment (the type assignment of $\Sigma$) is concerned. The formula $\bigwedge_{\psi \in D} \psi \wedge \phi'$ in this theorem is syntactically first-order. Denote this formula by $\gamma$. Our next step is to define a model-preserving translation from syntactically first-order formulas to first-order formulas.

To make $\gamma$ into a first-order formula, we should get rid of $\true$ and $\false$ occurring in a formula context. To preserve the semantics, we should also add axioms for the boolean sort, since in first-order logic all sorts are uninterpreted, while in \folb\ the interpretations of the boolean sort and constants $\true$ and $\false$ are fixed. 

To fix the problem, we will add axioms expressing that the boolean sort has two elements and that $\true$ and $\false$ represent the two distinct elements of this sort.
\begin{equation}\label{axiom:bool}
  \forall (x:\bool)(x \eql \true \vee x \eql \false) \wedge \true \not\eql \false.
\end{equation}
Note that this formula is a tautology in \folb, but not in FOL.

Given a syntactically first-order formula $\gamma$, we denote by $\toFOL{\gamma}$ the formula obtained from $\gamma$ by replacing all occurrences of $\true$ and $\false$ in a formula context by logical constants $\top$ and $\bot$ (interpreted as always true and always false), respectively and adding formula \eqref{axiom:bool}.

\begin{theorem}\label{thm:model:preservation:2}\rm
  Let $\context$ is a type assignment and $\gamma$ be a syntactically first-order formula such that $\context \vdash \gamma:\bool$.
  \begin{enumerate}
  \item Suppose that $\intI$ is a $\context$-interpretation and $\intI \models \gamma$ in \folb. Then $\intI \models \toFOL{\gamma}$ in first-order logic.
  \item Suppose that $\intI$ is a $\context$-interpretation and $\intI \models \toFOL{\gamma}$ in first-order logic. Consider the \folb-interpretation $\intI'$ that is obtained from $\intI$ by changing the interpretation of the boolean sort $\bool$ by $\{0,1\}$ and the interpretations of $\true$ and $\false$ by the elements $1$ and $0$, respectively, of this sort. Then $\intI' \models \gamma$ in \folb. \QED
  \end{enumerate}
\end{theorem}

Theorems~\ref{thm:model:preservation} and~\ref{thm:model:preservation:2} show that our translation preserves models. Every model of the original formula can be extended to a model of the translated formulas by adding values of the function symbols introduced during the translation. Likewise, any first-order model of the translated formula becomes a model of the original formula after changing the interpretation of the boolean sort to coincide with its interpretation in \folb.

\section{Superposition for \folb{}}
\label{sec:superposition}

In 
Section~\ref{sec:folb-to-fol} we presented a model-preserving
syntactic translation of \folb{} to FOL. 
Based on this translation, automated reasoning about \folb{} formulas
can be done by translating a \folb{} formula into a FOL
formula, and using an automated first-order theorem prover on the resulting FOL formula. 
State-of-the-art first-order theorem provers, such as Vampire~\cite{Vampire13}, E~\cite{E13} and
Spass~\cite{Spass}, implement superposition calculus for proving first-order formulas. Naturally, we would like to have a translation exploiting such provers in an efficient manner. 

Note however that our translation adds the two-element domain axiom 
$\forall (x:\bool)\allowbreak(x \eql \true \vee x \eql \false)$ for the boolean sort. This axioms will be converted to the clause
\begin{equation}\label{clause:T|F}
  x \eql \true \vee x \eql \false,
\end{equation}
where $x$ is a boolean variable. In this section we
explain why this axiom requires a special treatment and propose a solution to overcome problems caused by its presence. 
%

We assume some basic understanding of first-order theorem proving and superposition calculus, see, e.g.~\cite{Ganzinger01,NieuwenhuisRubio:HandbookAR:paramodulation:2001}. We fix a superposition inference system for first-order logic with equality, parametrised by a simplification ordering $\succ$ on literals and a well-behaved literal selection function \cite{Vampire13}, that is a function that guarantees completeness of the calculus. We denote selected literals by underlining them. We assume that equality literals are treated by a dedicated inference rule, namely, the ordered paramodulation rule~\cite{Robinson1969}:
\begin{prooftree}
  \AxiomC{$\underline{l \eql r} \vee C$} \RightLabel{\quad$\text{if}\ \theta = \mathrm{mgu}(l, s)$,}
  \AxiomC{$\underline{L[s]} \vee D$}
  \BinaryInfC{$(L[r] \vee C \vee D)\theta$}
\end{prooftree}
where $C,D$ are clauses, $L$ is a literal, $l,r,s$ are terms, $\mathrm{mgu}(l, s)$ is a most general unifier of $l$ and $s$, and $r\theta \not\succeq l\theta$.
The notation $L[s]$ denotes that $s$ is a subterm of $L$, then $L[r]$ denotes the result of replacement of $s$ by $r$.

Suppose now that we use an off-the-shelf superposition theorem prover to reason about FOL formulas obtained by our translation. W.l.o.g, we assume that $\true \succ \false$ in the term ordering used by the prover. Then self-paramodulation (from $\true$ to $\true$) can be applied to clause~\eqref{clause:T|F} as follows:
\begin{prooftree}
  \AxiomC{$\underline{x \eql \true} \vee x \eql \false$}
  \AxiomC{$\underline{y \eql \true} \vee y \eql \false$}
  \BinaryInfC{$x \eql y \vee x \eql \false \vee y \eql \false$}
\end{prooftree}
The derived clause $x \eql y \vee x \eql false \vee y \eql \false$ is a recipe for disaster, since the literal $x \eql y$ must be selected and can be used for paramodulation into every non-variable term of a boolean sort. Very soon the search space will contain many clauses obtained as logical consequences of clause \eqref{clause:T|F} and results of paramodulation from variables applied to them. This will cause a rapid degradation of performance of superposition-based provers.

To get around this problem, we propose the following solution. First, we will choose term
orderings $\succ$ having the following properties: $\true\succ\false$ and $\true$ and
$\false$ are the smallest ground terms w.r.t.\ $\succ$. Consider now all ground instances of \eqref{clause:T|F}. They have the form $s \eql \true \vee s \eql \false$, where $s$ is a ground term. When $s$ is either $\true$ or $\false$, this instance is a tautology, and hence redundant. Therefore, we should only consider instances for which $s \succ \true$. This prevents self-paramodulation of \eqref{clause:T|F}.

Now the only possible inferences with \eqref{clause:T|F} are inferences of the form
\[
\infer[,]{C[\true] \vee s \eql \false}%
{\underline{x \eql \true} \vee x \eql \false & C[s]}
\]
where $s$ is a non-variable term of the sort $\bool$.
To implement this, we can remove clause \eqref{clause:T|F} and add as an extra inference rule to the superposition calculus the following rule:
\[
\infer[,]{C[\true] \vee s \eql \false}%
{C[s]}
\]
where $s$ is a non-variable term of the sort $\bool$.

\section{TPTP support for \folb{}}
\label{sec:tptp}

The typed monomorphic first-order formulas subset, called TFF0, of the TPTP language~\cite{sutcliffe2012tptp}, is a representation language for many-sorted first-order logic. It contains \verb'if'-\verb'then'-\verb'else' and \verb'let'-\verb'in' constructs (see below), which is useful for applications, but is inconsistent in its treatment of the boolean sort. It has a predefined atomic sort symbol \verb'$o' denoting the boolean sort. However, unlike all other sort symbols, \verb|$o| can only be used to declare the return type of predicate symbols. This means that one cannot define a function having a boolean argument, use boolean variables or equality between booleans. 

Such an inconsistent use of the boolean sort results in having two kinds of \verb'if'-\verb'then'-\verb'else' expressions and four kinds of \verb'let'-\verb'in' expressions. For example, a \folb-term $\letin{f(x_1:\sigma_1, \ldots, x_n:\sigma_n)}{s}{t}$ can be represented using one of the four TPTP alternatives \verb|$let_tt|, \verb|$let_tf|, \verb|$let_ft| and \verb|$let_ff|, depending on whether $s$ and $t$ are terms or formulas. 

Since the boolean type is second-class in TPTP, one cannot directly represent formulas coming from program analysis and interactive theorem provers, such as formulas \eqref{formula:contains} and \eqref{formula:subset-sorted} of Section~\ref{sec:introduction}.

We propose to modify the TFF0 language of TPTP to coincide with \folb. It is not late to do so, since there is no general support for \verb'if'-\verb'then'-\verb'else' and \verb'let'-\verb'in'. To the best of our knowledge, Vampire is currently the only theorem prover supporting full TFF0. Note that such a modification of TPTP would make multiple forms of \verb'if'-\verb'then'-\verb'else' and \verb'let'-\verb'in' redundant. It will also make it possible to directly represent the SMT-LIB core theory.

We note that our changes and modifications on TFF0 can also be applied to the TFF1 language of TPTP~\cite{blanchette2013tff1}. TFF1 is  a polymorphic extension of TFF0 and its formalisation  does not treat the boolean sort. Extending our work to TFF1 should not be hard but has to be done in detail.

\section{Related work}
\label{sec:related}

Handling boolean terms as formulas is common in the SMT community. The SMT-LIB project~\cite{SMT-LIB} defines its core logic as first-order logic extended with the distinguished first-class boolean sort and the \verb'let'-\verb'in' expression used for local bindings of variables. The core theory of SMT-LIB defines logical connectives as boolean functions and the ad-hoc polymorphic \verb'if'-\verb'then'-\verb'else' ($ite$) function, used for conditional expressions. 
The language \folb\ defined here extends the SMT-LIB core language with local function definitions,
using \verb'let'-\verb'in' expressions defining functions of arbitrary, and not just zero, arity. This, \folb\ contains both this language and the TFF0 subset of TPTP. Further, we present a translation of \folb\ to FOL and show how one can improve superposition theorem provers to reason with the boolean sort. 


Efficient superposition theorem proving in finite domains, such as the boolean domain, is also discussed in~\cite{HillenbrandWeidenbach13}. The approach of~\cite{HillenbrandWeidenbach13} sometimes falls back to enumerating instances of a clause by instantiating finite domain variables with all elements of the corresponding domains. We point out here that for the boolean (i.e., two-element) domain there is a simpler solution. However, the approach of~\cite{HillenbrandWeidenbach13} also allows one to handle domains with more than two elements. One can also generalise our approach to arbitrary finite domains by using binary encodings of finite domains, however, this will necessarily result in loss of efficiency, since a single variable over a domain with $2^k$ elements will become $k$ variables in our approach, and similarly for function arguments.

\section{Conclusion}
\label{sec:conclusions}

We defined first-order logic with the first class boolean sort (\folb{}). It extends ordinary many-sorted first-order logic (FOL) with (i) the boolean sort such that terms of this sort are indistinguishable from formulas and (ii) \verb'if'-\verb'then'-\verb'else' and \verb'let'-\verb'in' expressions. The semantics of \verb'let'-\verb'in' expressions in \folb{} is essentially their semantics in functional programming languages, when they are not used for recursive definitions. In particular, non-recursive local functions can be defined and function symbols can be bound to a different sort in nested \verb'let'-\verb'in' expressions.

We argued that these extensions are useful in reasoning about problems coming from program analysis and interactive theorem proving. The extraction of properties from certain program definitions (especially in functional programming languages) into \folb{} formulas is more straightforward than into ordinary FOL formulas and potentially more efficient. In a similar way, a more straightforward translation of certain higher-order formulas into \folb{} can facilitate proof automation in interactive theorem provers.

\folb{} is a modification of FOL and reasoning in it reduces to reasoning in FOL. We gave a translation of \folb{} to FOL that can be used for proving theorems in \folb{} in a first-order theorem prover. We further discussed a modification of superposition calculus that can reason efficiently in presence of the boolean sort. Finally, we pointed out that the TPTP language can be changed to support \folb{}, which will also simplify some parts of the TPTP syntax.

Implementation of theorem proving support for \folb{}, including its super\-po\-sition-friendly translation to CNF, is an important task for future work. Further, we are also interested in extending \folb{} with theories, such as the theory of integer linear arithmetic and arrays.


\bibliographystyle{splncs03}
\bibliography{refs}



\end{document}